\documentclass[10pt,letterpaper]{article}
\usepackage[margin=1in]{geometry}
\usepackage[T1]{fontenc}
\usepackage[utf8]{inputenc}
\usepackage{lmodern}
\usepackage{amsmath,amssymb,booktabs,graphicx,array}
\usepackage{microtype}
\usepackage{hyperref}
\usepackage{caption}
\usepackage{float}
\hypersetup{hidelinks,pdftitle={When AI Gets Cheaper, Do Prices Follow? Information Delays and Algorithmic Pricing},pdfauthor={Krish Tadigotla}}
\title{When AI Gets Cheaper, Do Prices Follow?\\[0.2em]\large Information Delays and Algorithmic Pricing}
\author{Krish Tadigotla\\\small Texas Academy of Mathematics and Science (TAMS)\\\small University of North Texas\\\small Computer Science\\\small \texttt{KrishTadigotla@my.unt.edu}}
\date{}
\begin{document}
\maketitle
\begin{abstract}
As AI models become cheaper to run, a central question for buyers and market designers is whether providers will lower their prices accordingly. The answer may depend on how automated pricing systems respond to competitors. A firm can observe outdated rival prices even when it can revise its own price frequently; another may have current information but be unable to reprice immediately. These are distinct constraints, and their combined effects on cost pass-through and consumer welfare are uncertain. We isolate these mechanisms in a controlled simulation to examine how information delays and repricing restrictions interact when production costs fall. In a differentiated duopoly with tabular Q-learning agents, a $2\times2\times2\times2$ factorial design crosses information age, repricing cadence, cost regime, and firm asymmetry. We train 60 paired market seeds in each of 16 conditions (960 market-condition runs) and evaluate frozen policies using expected logit consumer surplus. In the symmetric declining-cost environment, the prespecified lag-by-cadence interaction is $-0.00025$ normalized welfare units, with a 95\% paired-seed bootstrap interval of $[-0.0402,\,0.0393]$. Neither this interaction nor the two prespecified moderation contrasts is statistically distinguishable from zero. In the fast-information, fast-repricing treatment, mean discrete price changes amount to 55.6\% and 44.0\% of the imposed cost reductions for the two firms. Welfare under learned policies is below the corresponding static grid-Nash benchmark in every treatment cell. The study provides a reproducible comparison of pricing-timing rules, but the findings are conditional on the simulated market and do not establish the pricing practices or collusion of actual AI providers.
\end{abstract}
\noindent\textbf{Keywords:} algorithmic pricing; Q-learning; information delay; cost pass-through; consumer welfare; computational industrial organization.\\
\textbf{JEL classification:} C63, D43, L13, L86.

\section{Introduction}
The price of large-language-model inference has fallen sharply, while available models and hosting arrangements increasingly differ in quality, cost, and specialization. Evidence from the OpenRouter inference marketplace documents sharp reductions in quality-adjusted prices and considerable entry \cite{demirer2026}; theoretical work describes the richer token-budget and menu choices facing LLM suppliers \cite{bergemann2026}. A basic industrial-organization question follows: when production becomes cheaper, how much of the saving reaches buyers?

Whether firms pass cost savings to consumers may depend on both how frequently they adjust prices and how quickly they observe competitors' prices. A firm's ability to reprice every period does not imply that it has a current rival-price feed. Conversely, a firm can observe today's rival price while being constrained to maintain yesterday's quote. Repricing frequency and information delay are separate constraints. Prior research has investigated automated price adjustment, strategic choice of update frequency, and the emergence of supracompetitive outcomes \cite{brown2023,calvano2020,conjeaud2026,klein2021}. Related work also studies Q-learning with stochastic production costs \cite{ballestero2026}. We therefore do not claim that either dynamic updating, imperfect information, or variable costs is novel in itself.

We test how delayed rival-price information interacts with synchronized repricing under constant or declining marginal costs. Firms are either symmetric or differ jointly in quality and cost. Outcomes are measured in expected logit buyer surplus as well as prices and profits. The design does not treat high prices or recurrent price paths as proof of autonomous collusion, an interpretation disputed in recent methodological work \cite{moraes2026,denboer2026}. The model permits us to attribute differences to programmed interventions within a synthetic market, not to actual provider conduct.

The primary interaction is not statistically distinguishable from zero; differences across individual treatment-cell means do not establish the proposed interaction. We also find average discrete price adjustment smaller than the assumed cost reduction in one highlighted treatment; however, only one cost-shock magnitude is imposed, so the data do not identify a general pass-through schedule. Results are sensitive to several learning assumptions, and the available data do not establish Q-value convergence during training. The interpretation is confined to the simulated environment.

\section{Related research and contribution}
Simulations by Calvano et al. \cite{calvano2020} show that independent Q-learning agents can generate high prices in a differentiated Bertrand market. Klein \cite{klein2021} studies sequential price setting, and Brown and MacKay \cite{brown2023} model differences in pricing technology and algorithmic responses in market equilibrium. Conjeaud et al. \cite{conjeaud2026} investigate asynchronous updates under Poisson clocks and different rival-price information interfaces, showing that synchronization and observability matter for simulated prices. Our design does not replicate their continuous-time Poisson environment: it independently fixes feed age and update cadence in discrete market ticks.

Recent contributions impose further caution. Ballestero \cite{ballestero2026} varies production costs under sequential learning; hence cost shocks alone cannot anchor novelty. den Boer et al. \cite{denboer2026} question whether learned high prices satisfy economically relevant collusion criteria. de Moraes and De-Losso \cite{moraes2026} illustrate why simple frozen-policy perturbations cannot unambiguously identify deterrence mechanisms. Bichler et al. \cite{bichler2026} find that agent-interface choices can alter price competition. We use these studies to motivate explicit timing controls, nonlearning benchmarks, direct consumer-welfare measurement, and restrained mechanism claims. We did not run deviation-response experiments.

The design has three methodological features. First, the price-feed lag and the timing of permissible quote revisions are independently varied and structurally distinguishable even when firms update simultaneously. Second, the analysis tests whether the timing interaction changes with an exogenous production-cost reduction and asymmetric service values/costs. Third, inference proceeds at the independent market-seed level and preserves all treatment cells, including results unfavorable to a dramatic hypothesis. The estimates apply to this simulated market. The comparison does not establish priority over all related work or describe actual providers' pricing practices.

\section{Environment and methods}
\subsection{Differentiated demand and welfare}
Two sellers $i\in\{A,B\}$ quote prices $p_{it}$ from a finite set $\mathcal{P}=\{0.60,0.75,\ldots,3.00\}$. With perceived quality $v_i$, logit scale $\mu=0.25$, and an outside good with normalized deterministic utility zero, their demand shares and expected buyer surplus are
\begin{equation}
 s_{it}=\frac{\exp[(v_i-p_{it})/\mu]}{1+\sum_{j\in\{A,B\}}\exp[(v_j-p_{jt})/\mu]},
\end{equation}
\begin{equation}
 s_{0t}=1-s_{At}-s_{Bt},\qquad CS_t=\mu\log\!\left(1+\sum_{j\in\{A,B\}}\exp[(v_j-p_{jt})/\mu]\right).
\end{equation}
Firm profit is $\pi_{it}=(p_{it}-c_{it})s_{it}$. The potential-buyer mass is one. Prices, costs, and welfare measures are expressed in normalized units per market tick, not dollars or billed tokens. The simulation uses numerical log-sum-exp for stability.

The symmetric treatment uses $v=(2.0,2.0)$ with initial cost $c_H=(1.00,1.00)$ and final cost $c_L=(0.70,0.70)$. The jointly asymmetric treatment uses $v=(2.2,1.9)$, $c_H=(1.00,0.78)$, and $c_L=(0.70,0.55)$. Because quality and cost vary together, their individual effects cannot be identified. This treatment also says nothing about model licensing. Flat-cost cells remain at $c_H$. Declining-cost cells undergo an exogenous shift to $c_L$ at $t=1000$ during frozen evaluation. During training, a geometric waiting time is drawn independently in each episode, avoiding a known deterministic shock date in learning.

\subsection{Separate information and price-revision clocks}
At tick $t$, each agent observes its own prevailing price and cost regime plus the rival's historically posted price $p_{-i,t-L}$. We vary the feed age $L\in\{1,4\}$ market ticks. An agent can revise its own price only when $t\equiv0\pmod F$, where $F\in\{1,3\}$; otherwise its quote is held. Both agents use simultaneous, synchronized eligibility schedules within the prespecified experiment. This isolates cadence from staggered updating; it does not examine unequal firm clocks.

The $L=1$ versus $L=4$ contrast ensures variation in observed rival history even with synchronized three-tick repricing. With synchronized revisions every three ticks, an age-one feed reflects the rival's last revision, while an age-four feed can omit that revision. With $F=3$, ages one and three would otherwise produce identical observed rival quotes. No contemporaneous rival action leaks into the agent's observation.

\subsection{Learning and equilibrium references}
Each firm has a tabular action-value table with observed state $S_{it}=(p_{it^-},p_{-i,t-L},k_t)$, where $k_t$ denotes the observable high/low cost regime. This is a partially observed state representation, not a claim that the opponent's strategy is fully Markov. Training uses $\alpha=0.10$, per-tick discount factor $\gamma=0.95$, and epsilon-greedy exploration declining from 1.00 to 0.02 over each agent's scheduled decisions. Profits accrue every tick. For a held price duration $d$, the duration-aware update is
\begin{equation}
 Q(S_t,a_t)\leftarrow(1-\alpha)Q(S_t,a_t)+\alpha\!\left[\sum_{j=0}^{d-1}\gamma^j\pi_{i,t+j}+\gamma^d\max_{a'}Q(S_{t+d},a')\right],
\end{equation}
with the continuation value removed at episode termination. Each market trains for 10 episodes of 15,000 ticks; Q tables persist while market prices and feed buffers reset at episode boundaries. Separate seeded streams govern agents and cost-shock waiting times.

We compute cost/configuration-specific finite-grid one-shot Nash prices and joint-profit-maximizing prices. The Nash profile satisfies
\begin{equation}
 \pi_i\bigl(p_i^N,p_{-i}^N;c_i\bigr)\geq\pi_i\bigl(p_i,p_{-i}^N;c_i\bigr)\quad\text{for every }p_i\in\mathcal{P}.
\end{equation}
The computed pure-grid equilibrium is unique in each required state. A separate continuous-price first-order-condition solver provides a numerical cross-check, but the grid Nash is the operative comparison. The static Nash is a one-period benchmark, not the equilibrium of the repeated stochastic game.

\subsection{Design, evaluation, and prespecified analysis}
The prespecified 16 cells cross $L\in\{1,4\}$, $F\in\{1,3\}$, cost type (flat/declining), and configuration (symmetric/asymmetric). Independently initialized market seed IDs 200000--200059 are reused across all conditions, yielding 60 paired markets per cell (960 market-condition training runs). Following 150,000 training ticks per market, Q is frozen and the model is evaluated for 2,000 ticks, starting from the high-cost static grid Nash. The frozen evaluation is a standardized reset experiment, not continuation from the final training price history. All 960 markets were retained, including unsuccessful or uncompetitive learned outcomes.

Our prespecified market-level outcome is mean consumer surplus over post-shock ticks 1050--1999, 950 ticks in total:
\begin{equation}
 Y_{s,L,F,C,A}=\frac{1}{950}\sum_{t=1050}^{1999}CS_{s,L,F,C,A,t}.
\end{equation}
Let $D_s(C,A)=[Y_{s,4,3,C,A}-Y_{s,1,3,C,A}]-[Y_{s,4,1,C,A}-Y_{s,1,1,C,A}]$. The prespecified contrasts are (H1) $\mathbb{E}[D_s(\mathrm{declining},\mathrm{symmetric})]$, (H2) the difference of H1 between declining and flat cost regimes, and (H3) the difference between asymmetric and symmetric firms in the declining-cost regime. These expectations refer only to the model's random initialization and training streams.

For each hypothesis, 95\% percentile confidence intervals use 10,000 resamples of whole paired market-seed blocks with a fixed bootstrap seed of 987654321. Secondary H2 and H3 paired-$t$ $p$-values receive Holm adjustment as a two-test family. A separate saturated $2^4$ descriptive regression includes market-seed fixed effects and standard errors clustered by the 60 independent seed blocks. The 1.92 million serially dependent tick observations are never interpreted as independent regression units.

\section{Results}
\subsection{Prespecified interactions: no rejection of the primary null}
\begin{table}[htbp]\centering\small
\begin{tabular}{lrrr}\toprule
Hypothesis & Estimate & 95\% paired bootstrap CI & $n$\\\midrule
H1: Lag $\times$ cadence & $-0.0003$ & $[-0.0402,+0.0393]$ & 60\\
H2: Additional cost moderation & $+0.0166$ & $[-0.0348,+0.0685]$ & 60\\
H3: Additional asymmetry moderation & $-0.0504$ & $[-0.1020,+0.0009]$ & 60\\\bottomrule
\end{tabular}
\caption{Prespecified seed-paired interaction estimates. Units: normalized logit buyer surplus per potential buyer per market tick. Percentile 95\% intervals resample 60 entire market seeds, 10,000 times. All intervals include zero. H2/H3 Holm-adjusted paired-$t$ $p$-values are 0.5312 and 0.1210 respectively.}
\end{table}
The H1 point estimate is $-0.0002533$ and the confidence interval ranges from $-0.04022$ to $+0.03929$. The estimate does not establish equivalence: the interval still admits meaningful effects in either direction on the model's normalized welfare scale. The secondary cost and asymmetry moderation intervals also include zero, and neither passes its adjusted inference criterion. The estimates do not establish whether outdated rival-price information strengthens or weakens the welfare effect of slower repricing.

\begin{figure}[htbp]\centering\includegraphics[width=\linewidth]{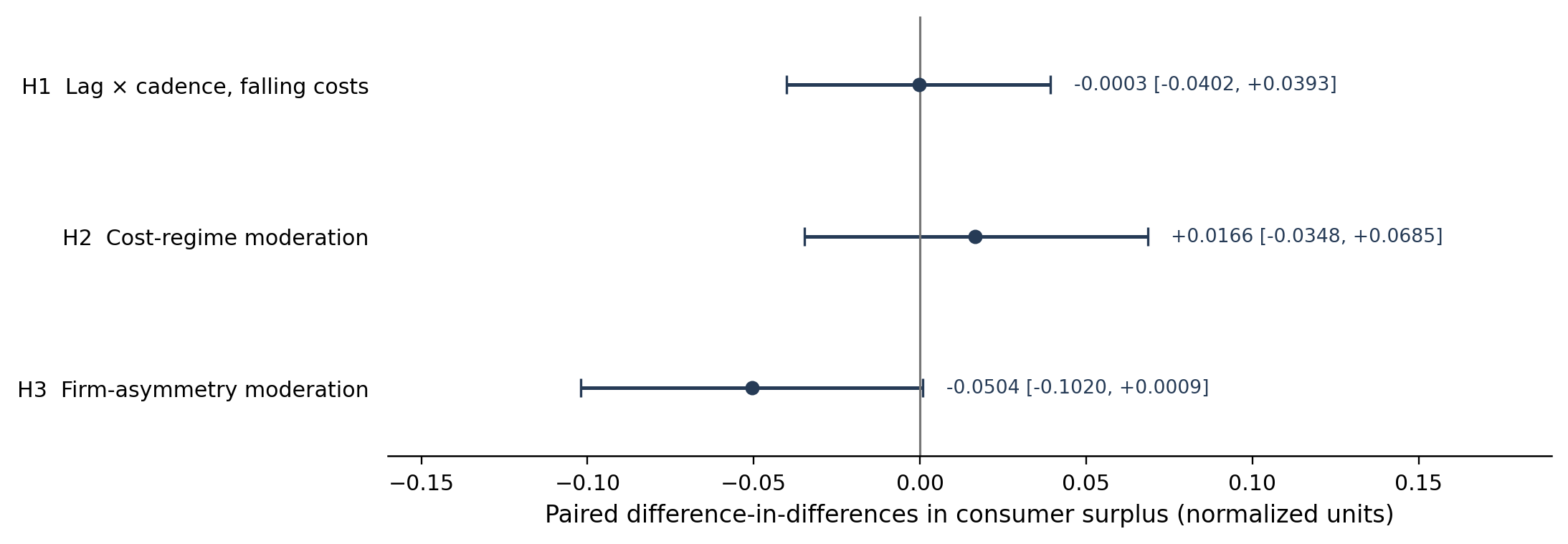}
\caption{Prespecified interaction contrasts. Points indicate paired-seed means and bars indicate 95\% percentile bootstrap intervals over 60 market seeds. The vertical reference line denotes zero interaction.}
\end{figure}

\subsection{Cell means and consumer welfare}
Figure 2 displays mean post-window welfare for all 16 cells and their whole-seed uncertainty. Differences between raw cell means exist, but individual pairwise differences do not substitute for the prespecified difference-in-differences test.
\begin{figure}[htbp]\centering\includegraphics[width=0.93\linewidth]{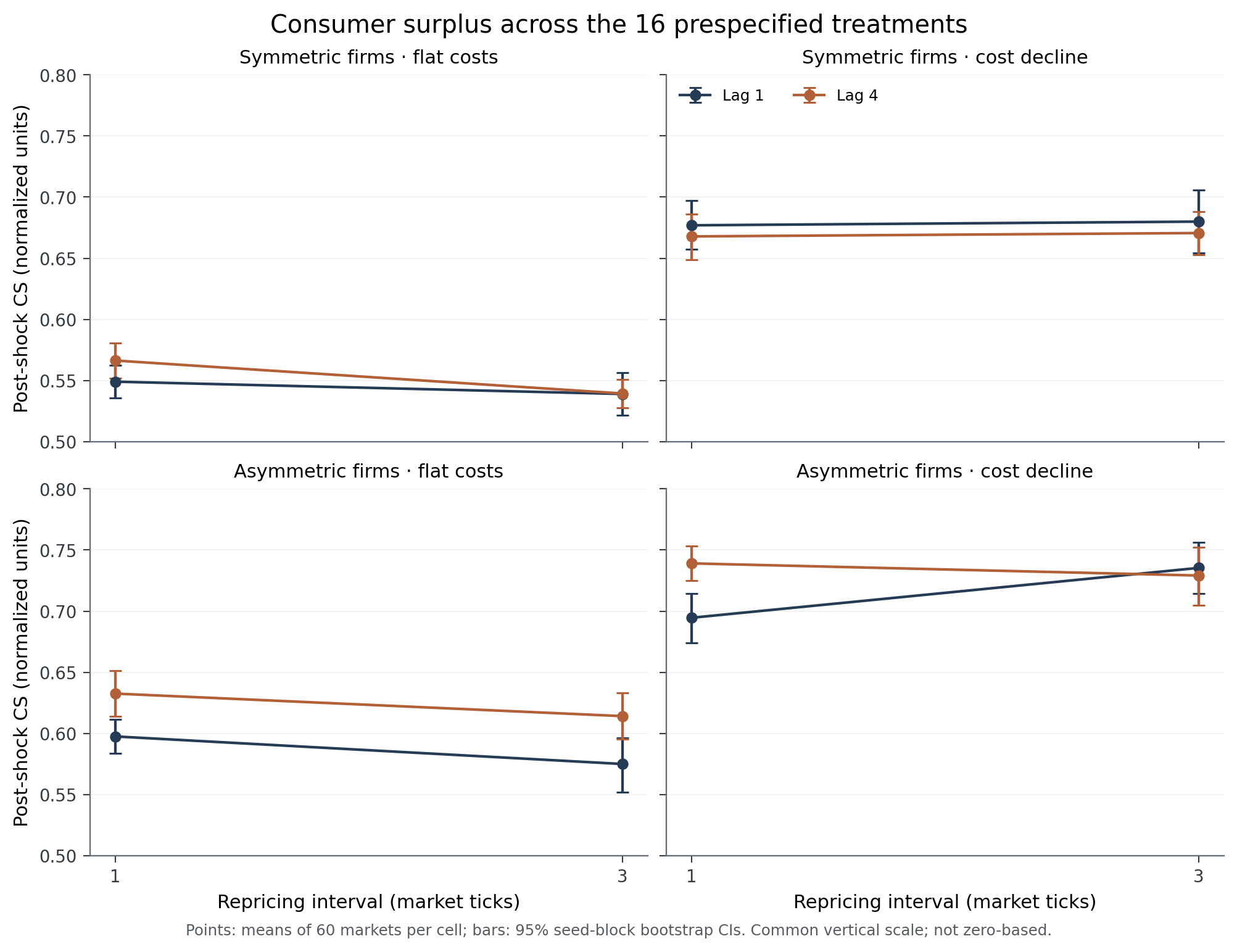}
\caption{Mean post-window logit consumer surplus across all 16 simulation conditions. Intervals are 95\% confidence intervals over market seeds; panels use comparable welfare scales.}
\end{figure}
For every learned-treatment cell mean, post-window welfare is below the appropriate current-cost one-shot grid Nash benchmark; mean gaps range from roughly $-0.333$ to $-0.123$ normalized units. This finding does not establish algorithmic deterrence, joint profit maximization, a stable learned equilibrium, or illegal collusion. The baseline itself is static and omits the adaptive strategies and partial observability of the repeated learning environment.

\subsection{Responses to cost reductions}
Every falling-cost cell has a negative mean change in posted price after the forced cost decline, but the price reductions are generally smaller than the imposed reductions in marginal cost. For the symmetric, $L=1$, $F=1$ case, the mean ratio of price change to cost change is 0.5559 for firm A and 0.4397 for firm B. The ratio uses prices averaged over ticks 800--999 versus 1050--1249, divided by the cost difference of $-0.30$. Confidence intervals based on market seeds are shown in Figure 4. There is only one shock magnitude per firm; a conventional continuous pass-through slope or empirical elasticity is not identified here.
\begin{figure}[htbp]\centering\includegraphics[width=0.96\linewidth]{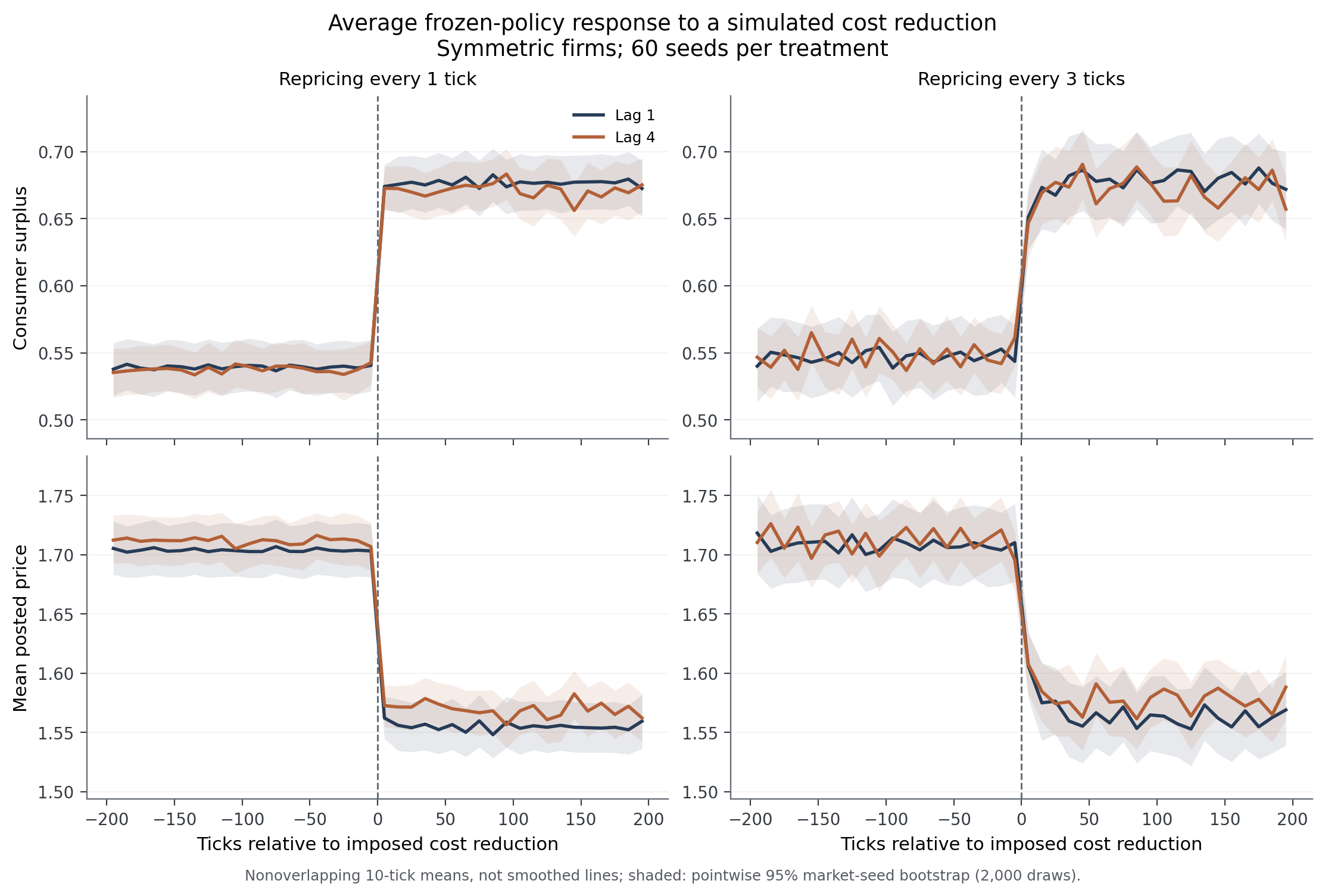}
\caption{Post-learning price and buyer-surplus response centered on the synthetically imposed cost decline at tick 1000 in the symmetric market. Trajectories are averaged into disclosed nonoverlapping 10-tick bins; bands are pointwise 95\% seed-bootstrap intervals, not simultaneous event-study bands.}
\end{figure}
\begin{figure}[htbp]\centering\includegraphics[width=0.96\linewidth]{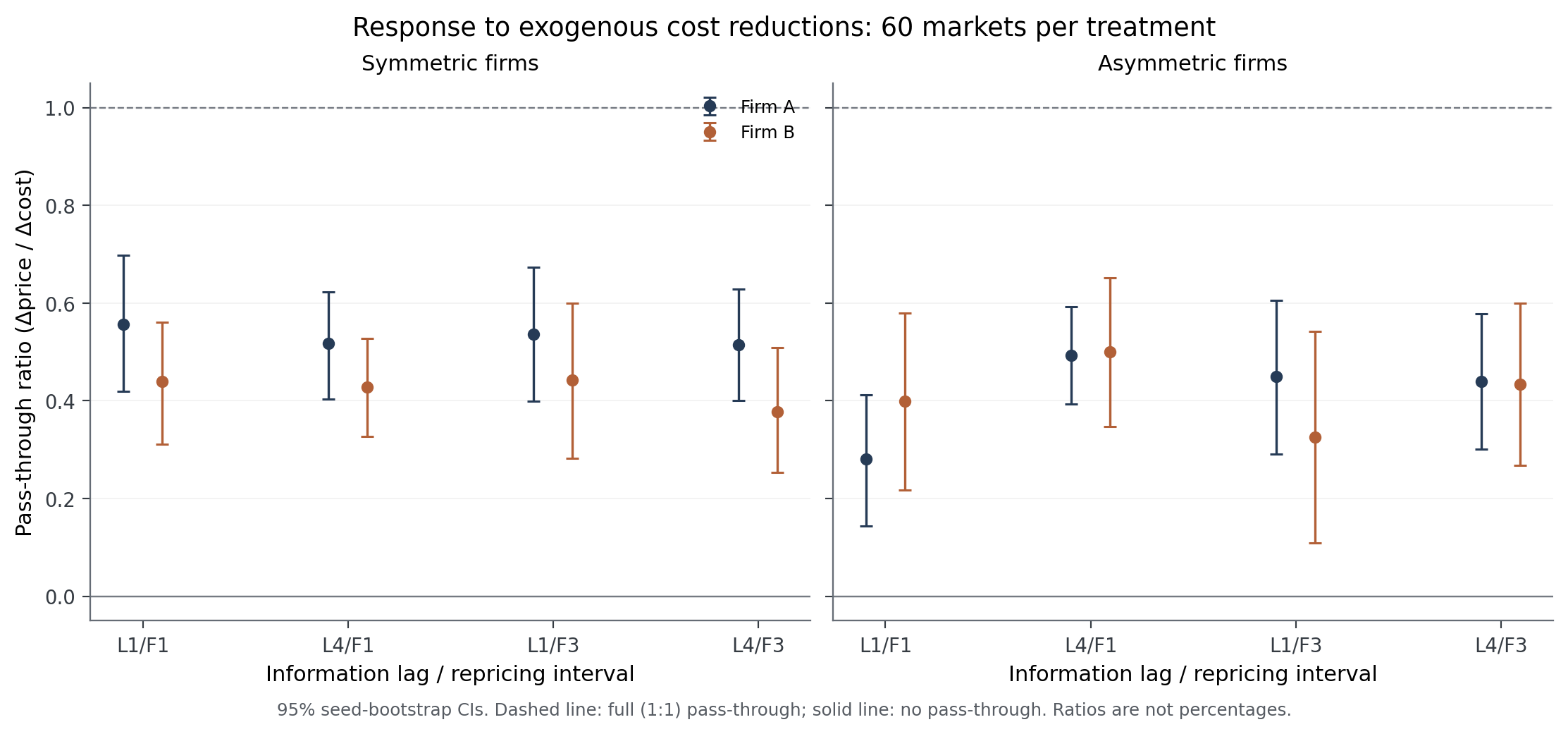}
\caption{Ratio of the mean simulated price change to the imposed marginal-cost change, by firm and treatment. Intervals resample market seeds; the reference value of one denotes full absolute pass-through.}
\end{figure}

\subsection{Benchmarks, learning sensitivity, and convergence}
As an implementation check, the constant-price control is invariant to the rival-price feed age. A deterministic myopic-best-response policy produces the corresponding grid-Nash reference in the saved control configurations, while the learned policies generally do not. The four prespecified nonconfirmatory learning-parameter variations yield timing-interaction estimates ranging from about $-0.068$ to $+0.040$, with exploratory intervals including zero; this does not exhaust parameter robustness. In a separate 10-seed longer-training comparison the exploratory interaction is about $+0.093$, but the intervention changes market time, learning decisions, and training exposure jointly.

The experiment does not establish convergence of training Q values. There are late price changes in frozen greedy evaluation in 882/960 market-condition runs; frozen cyclic policies can do this even if Q has converged, while a stable price path alone would also not prove Q convergence. The archived experiment lacks sequential training Q-table snapshots necessary for a direct convergence analysis. Accordingly, equilibrium and deterrence mechanisms remain unverified. Figure 5 and the appendix report all robustness outcomes rather than selecting favorable scenarios.
\begin{figure}[H]\centering\includegraphics[width=0.9\linewidth]{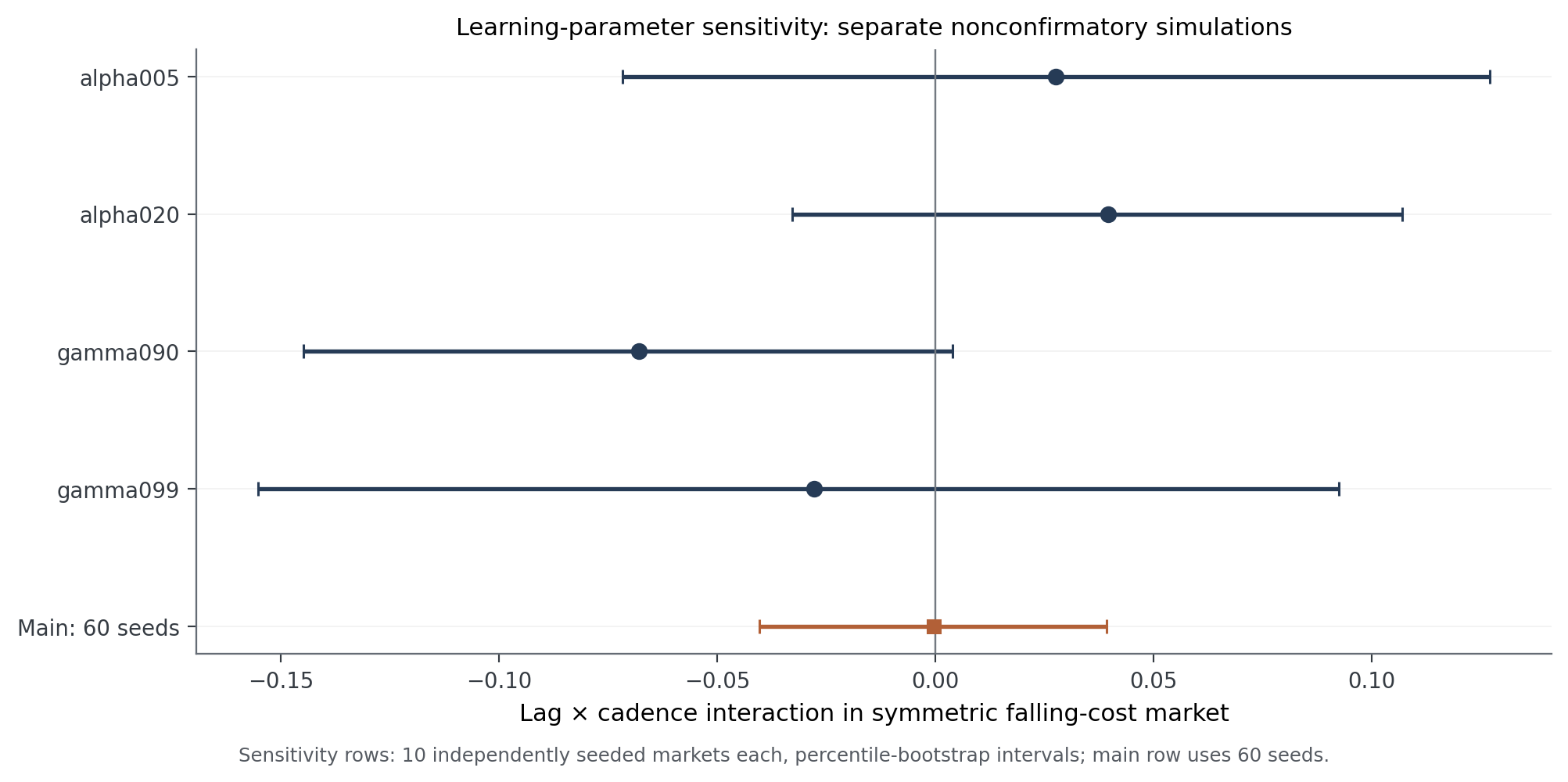}
\caption{Nonconfirmatory lag-by-cadence contrasts under changed learning rates or per-tick discount factors (10 seeds per setting), alongside the prespecified 60-seed baseline. All intervals cross zero; this panel is exploratory, not additional primary evidence.}
\end{figure}

\section{Discussion and limits}
Slower rival-price feeds can plausibly weaken rapid undercutting, but they can also impede responses that otherwise sustain higher prices. Neither sign is guaranteed in a partially observed repeated game. The estimated interaction is too imprecise to distinguish between these mechanisms. Interpreting an intervention therefore requires knowing which timing rule changes, how welfare is measured, which evaluation baseline is used, and how many learning opportunities the agents receive.

External validity is limited in several specific ways. The model has two firms with a single scalar price per firm, whereas actual LLM pricing often uses input/output token rates, budgets, volume discounts, context windows and model-specific quality. Demand elasticities, prices, marginal costs and timing distributions were stipulated rather than fit to market microdata. The asymmetric treatment jointly alters costs and consumer valuations. The agents are tabular Q-learners, not documented corporate pricing algorithms; real providers are not known to use this policy. Both firms update synchronously in every confirmatory cell. The discounted learning state is partially observed, and no global convergence or equilibrium existence result is demonstrated for the stochastic game.

A nonsignificant result does not imply that the effect is exactly zero. The confidence interval for H1 is nontrivial on the welfare scale, and the sensitivity analysis covers only a small subset of learning choices. Future studies should prespecify finer price grids, alternative demand scales, nonsynchronous update offsets, quality-only/cost-only asymmetry, and richer agents; retain training-time Q snapshots; and obtain price/quality observations from actual markets for external calibration. These extensions would require a separate design and independent data generation.

\section{Conclusion}
In this prespecified inference-market simulation, the interaction between rival-price feed age and repricing cadence is not statistically distinguishable from zero for post-cost-shock consumer welfare in the primary symmetric condition. Average discrete pass-through of the modeled production-cost reduction is incomplete in a highlighted treatment, and learned-policy mean welfare lies below static grid-Nash references, but the evidence does not prove strategic collusion, equilibrium convergence, or effects in observed LLM provider markets. The study provides a reproducible test of this timing interaction and a baseline for richer models of adaptive pricing. More comprehensive learning-convergence and external-validity analysis is needed before economic-policy or provider-level claims would be appropriate.

\appendix
\section{Full treatment results}
\begin{table}[htbp]\centering\small
\begin{tabular}{llrrrrr}\toprule
Firms & Costs & $F$ & $L$ & Mean CS & SD(CS) & Joint profit\\\midrule
Sym & Flat & 1 & 1 & 0.549 & 0.053 & 0.551\\
Sym & Flat & 1 & 4 & 0.566 & 0.059 & 0.542\\
Sym & Flat & 3 & 1 & 0.539 & 0.070 & 0.551\\
Sym & Flat & 3 & 4 & 0.539 & 0.047 & 0.548\\
Sym & Decl & 1 & 1 & 0.677 & 0.080 & 0.716\\
Sym & Decl & 1 & 4 & 0.668 & 0.074 & 0.718\\
Sym & Decl & 3 & 1 & 0.680 & 0.103 & 0.702\\
Sym & Decl & 3 & 4 & 0.671 & 0.071 & 0.707\\
Asy & Flat & 1 & 1 & 0.597 & 0.056 & 0.658\\
Asy & Flat & 1 & 4 & 0.633 & 0.074 & 0.627\\
Asy & Flat & 3 & 1 & 0.575 & 0.088 & 0.664\\
Asy & Flat & 3 & 4 & 0.614 & 0.075 & 0.630\\
Asy & Decl & 1 & 1 & 0.695 & 0.082 & 0.818\\
Asy & Decl & 1 & 4 & 0.739 & 0.055 & 0.779\\
Asy & Decl & 3 & 1 & 0.735 & 0.083 & 0.779\\
Asy & Decl & 3 & 4 & 0.729 & 0.094 & 0.772\\\bottomrule
\end{tabular}
\caption{All 16 prespecified treatment cells (60 markets each). Mean and across-seed SD of post-window expected logit buyer surplus (CS) are in normalized units; joint profit is the sum of both sellers' mean profits. Sym. = symmetric firm configuration; Asy. = quality/cost asymmetric. $L$ is competitor-feed age in ticks; $F$ is quote-revision interval. No cell was dropped.}
\end{table}

\section{Reproduction of the statistical analysis}
The analysis reads the archived 960 market-condition simulations (60 paired random seeds in each of 16 cells). Each frozen-policy evaluation contains 2,000 simulated market ticks. The post-shock estimand averages ticks 1050--1999, while the discrete cost-pass-through calculation compares the specified pre- and post-shock windows in Section 4.3. Market-level H1--H3 contrasts are computed before inference; 95\% percentile confidence intervals resample all 60 seed blocks with 10,000 bootstrap draws and pseudorandom seed 987654321. No tick-level observations enter the bootstrap as independent units. The replication materials specify the Python environment, training protocol, data validation checks, and commands to regenerate tables and figures.

\section{Supplementary figures}
\begin{figure}[htbp]\centering\includegraphics[width=0.91\linewidth]{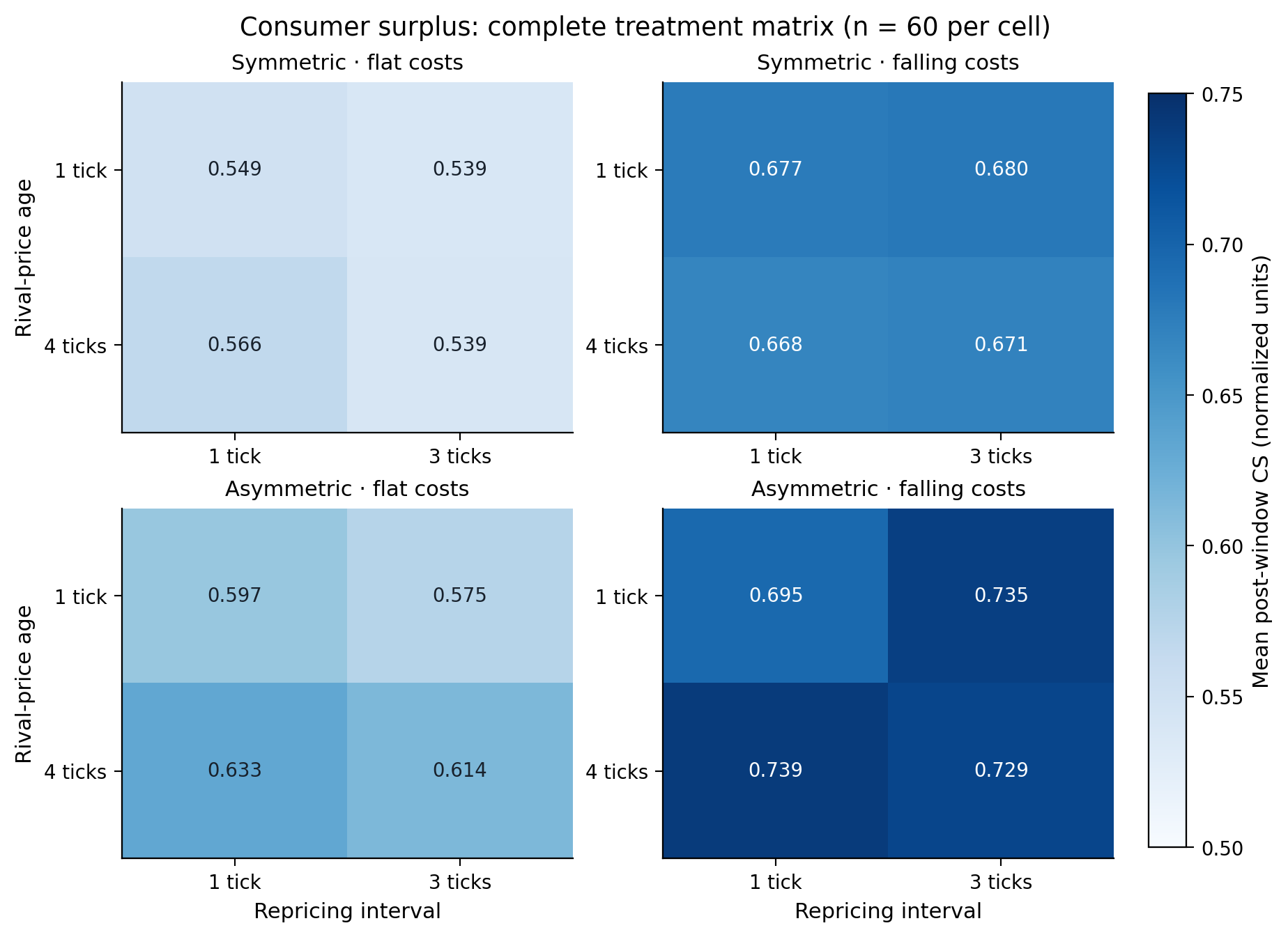}
\caption{Mean simulated consumer surplus for the 16 factorial cells. A common color scale is used across panels; color differences are descriptive rather than inferential.}
\end{figure}
\begin{figure}[htbp]\centering\includegraphics[width=0.91\linewidth]{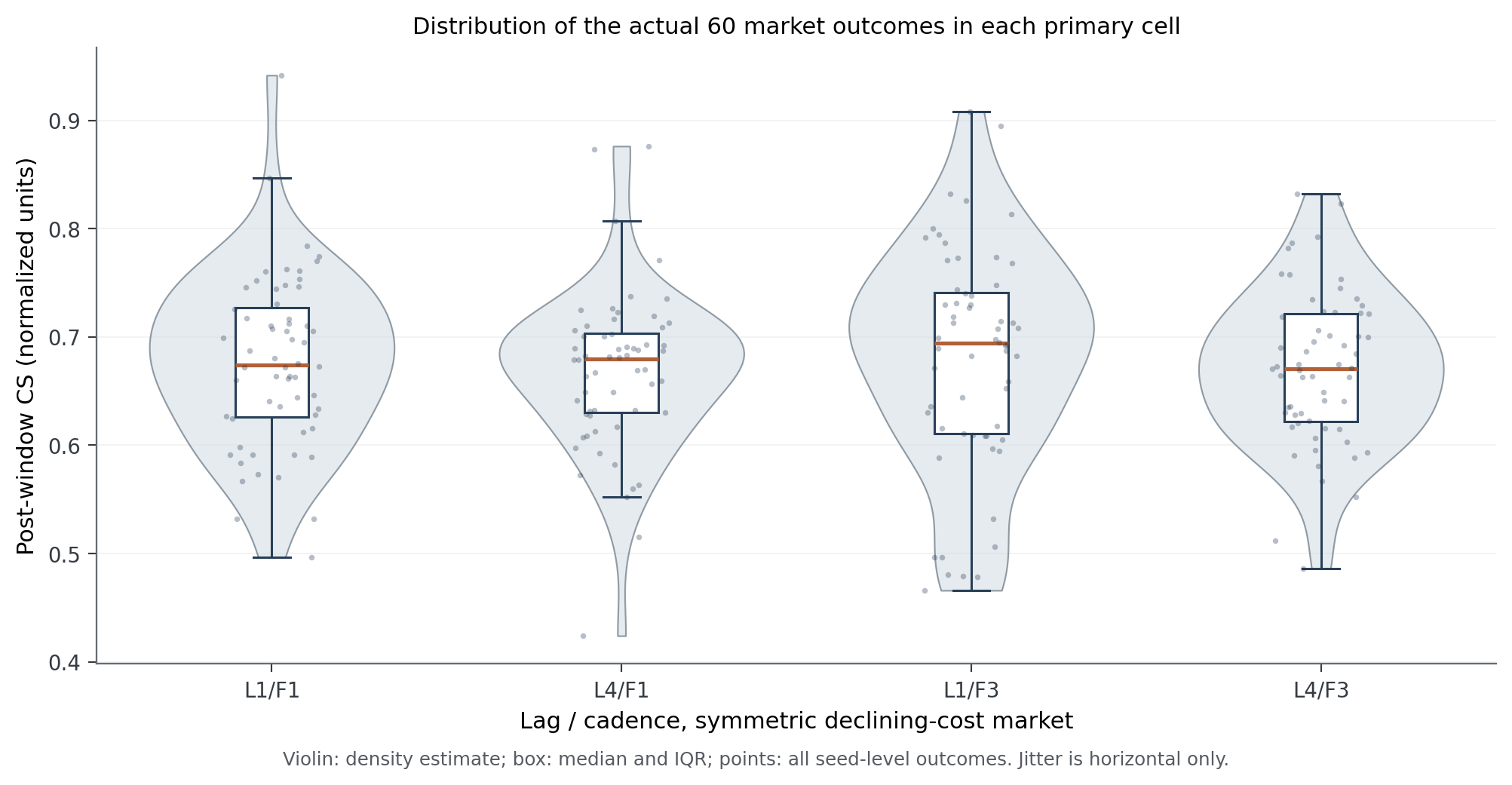}
\caption{Distribution of seed-level post-window consumer surplus for the four symmetric declining-cost cells ($n=60$ per cell). Points show individual simulated markets.}
\end{figure}
\begin{figure}[htbp]\centering\includegraphics[width=0.91\linewidth]{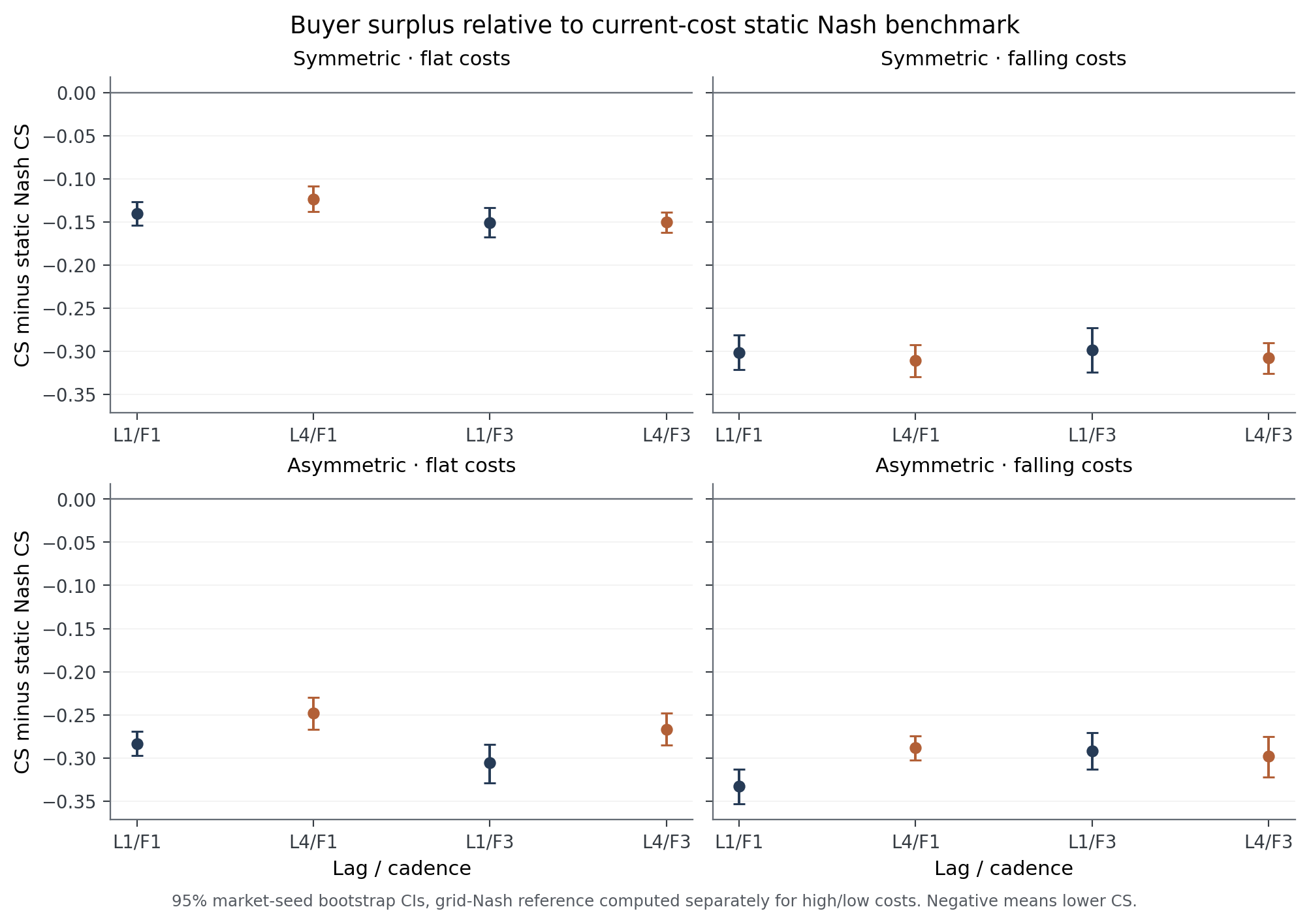}
\caption{Learned-policy consumer surplus less current-cost static grid-Nash consumer surplus, with uncertainty over market seeds. The one-period Nash solution is a descriptive benchmark.}
\end{figure}
\begin{figure}[htbp]\centering\includegraphics[width=0.91\linewidth]{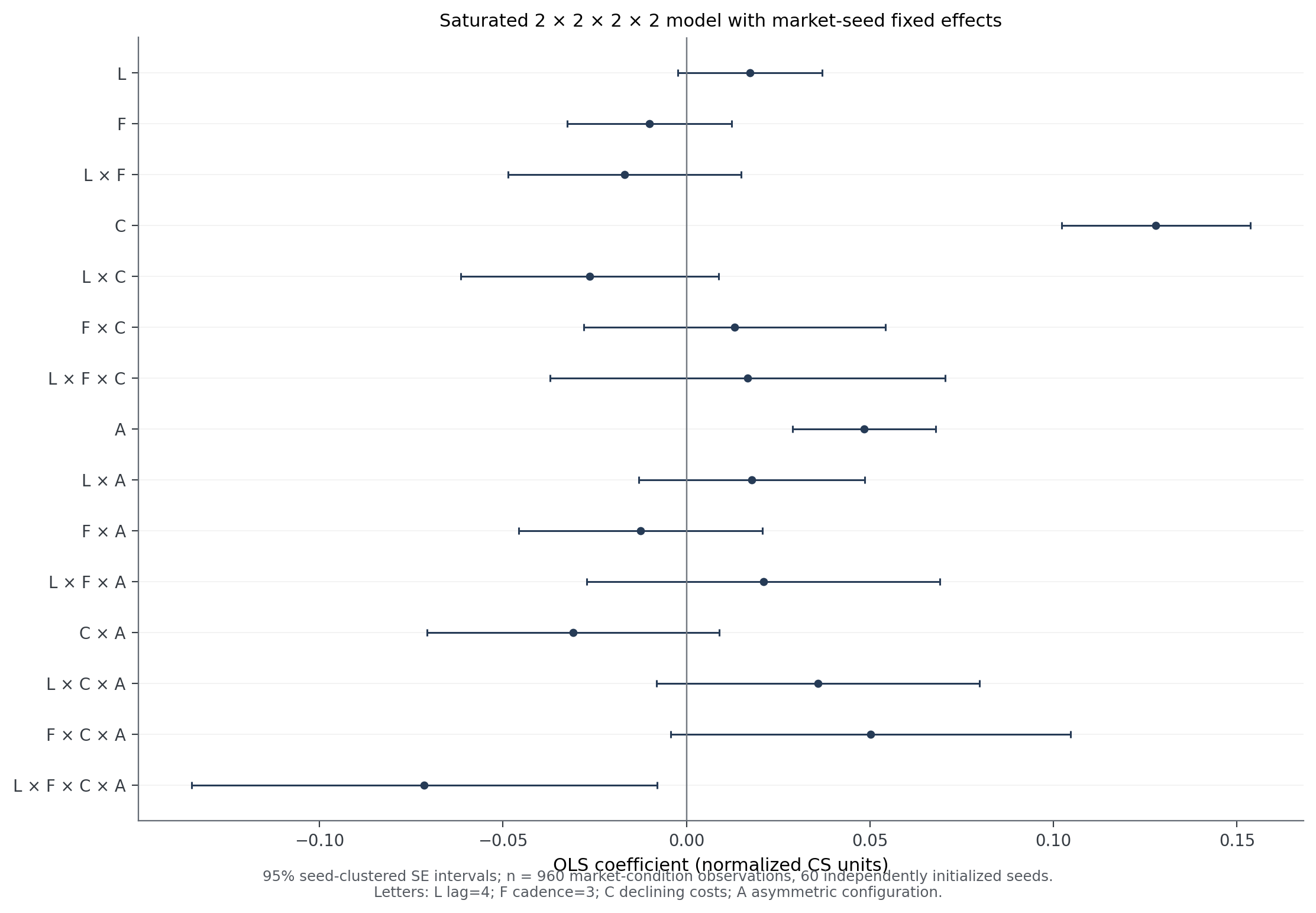}
\caption{Coefficients from the saturated factorial regression with seed fixed effects and standard errors clustered by seed. This is a secondary descriptive specification.}
\end{figure}
\begin{figure}[htbp]\centering\includegraphics[width=0.91\linewidth]{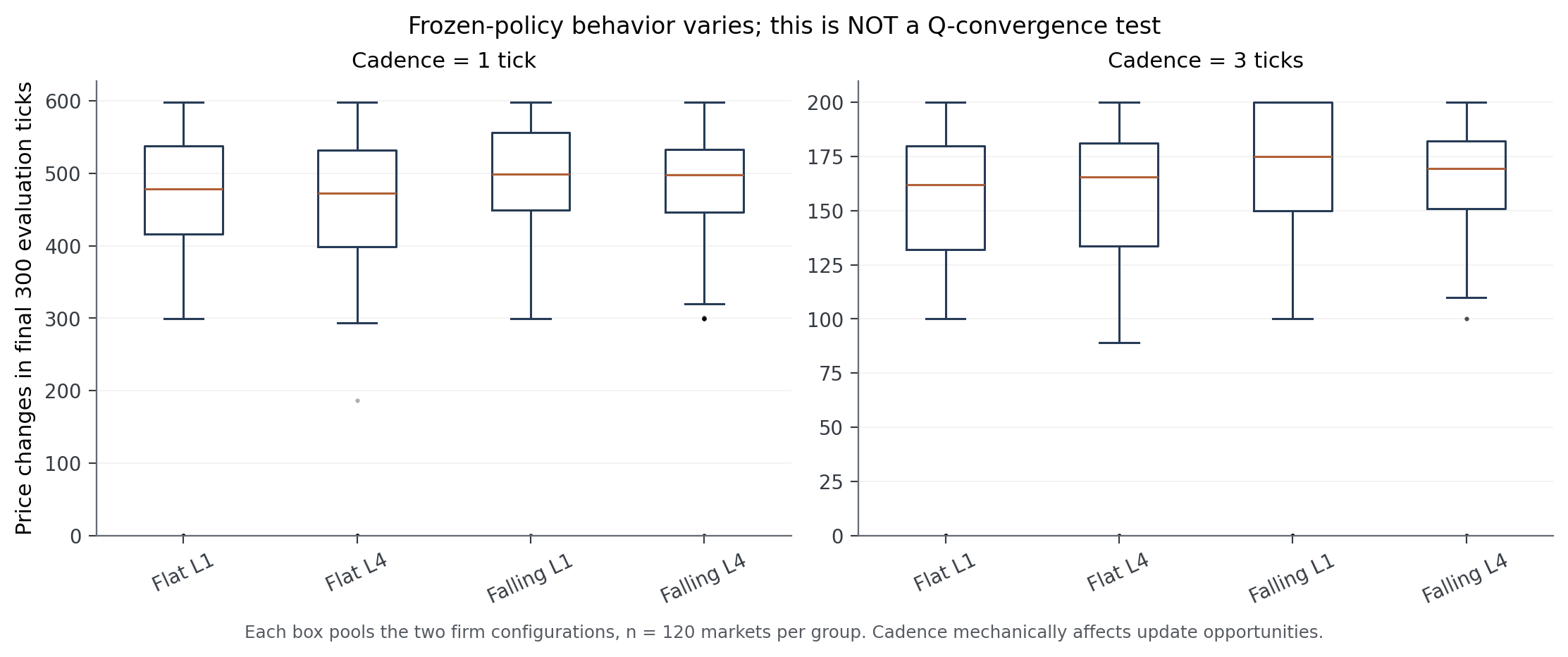}
\caption{Frequency of price changes and dispersion in the last 300 frozen-policy evaluation ticks. These diagnostics do not establish convergence of training Q values.}
\end{figure}

\begin{thebibliography}{10}
\bibitem{ballestero2026} Gonzalo Ballestero. Algorithmic collusion under sequential pricing and stochastic costs. \emph{International Journal of Industrial Organization}, 106:103281, 2026. doi: \href{https://doi.org/10.1016/j.ijindorg.2026.103281}{10.1016/j.ijindorg.2026.103281}.
\bibitem{bergemann2026} Dirk Bergemann, Alessandro Bonatti, and Alex Smolin. Menu pricing of large language models. Cowles Foundation Discussion Paper 2502, Yale Cowles Foundation, 2026. URL: \url{https://cowles.yale.edu/research/cfdp-2502-menu-pricing-large-language-models}.
\bibitem{bichler2026} Martin Bichler, Philip Hierhager, Kassian K\"ock, and Fabian Pieroth. How language model agents reshape price competition, 2026. URL: \url{https://arxiv.org/abs/2610.21711}.
\bibitem{brown2023} Zach Y. Brown and Alexander MacKay. Competition in pricing algorithms. \emph{American Economic Journal: Microeconomics}, 15(2):109--156, 2023. doi: \href{https://doi.org/10.1257/mic.20210158}{10.1257/mic.20210158}.
\bibitem{calvano2020} Emilio Calvano, Giacomo Calzolari, Vincenzo Denicol\`o, and Sergio Pastorello. Artificial intelligence, algorithmic pricing, and collusion. \emph{American Economic Review}, 110(10):3267--3297, 2020. doi: \href{https://doi.org/10.1257/aer.20190623}{10.1257/aer.20190623}.
\bibitem{conjeaud2026} Ivan Conjeaud, Gaspard Abel, and Argyris Kalogeratos. Algorithmic collusion under asynchronous price updating, 2026. URL: \url{https://arxiv.org/abs/2608.01406}.
\bibitem{moraes2026} Kau\^e Lopes de Moraes and Rodrigo De-Losso. Algorithmic collusion without deterrence: Auditing the punishment mechanism in q-learning pricing agents, 2026. URL: \url{https://arxiv.org/abs/2610.21867}.
\bibitem{demirer2026} Mert Demirer, Andrey Fradkin, and Nadav Tadelis. The emerging market for intelligence: How firms buy and sell AI. \emph{Journal of Economic Perspectives}, 40(3):23--46, 2026. doi: \href{https://doi.org/10.1257/jep.20261506}{10.1257/jep.20261506}.
\bibitem{denboer2026} Arnoud den Boer, Janusz M. Meylahn, and Maarten Pieter Schinkel. Artificial collusion: Examining supracompetitive pricing by q-learning algorithms. \emph{Management Science}, 2026. doi: \href{https://doi.org/10.1287/mnsc.2024.08557}{10.1287/mnsc.2024.08557}.
\bibitem{klein2021} Timo Klein. Autonomous algorithmic collusion: Q-learning under sequential pricing. \emph{RAND Journal of Economics}, 52(3):538--558, 2021. doi: \href{https://doi.org/10.1111/1756-2171.12383}{10.1111/1756-2171.12383}.
\end{thebibliography}
\end{document}